\documentclass[%
aip,
reprint,
]{revtex4-2}

\usepackage{graphicx}
\graphicspath{{/}}

\usepackage{amsmath}

\begin{document}


\title{Improved Filters for Angular Filter Refractometry} 



\author{P. V. Heuer}
\email[]{pheu@lle.rochester.edu}
\affiliation{University of Rochester Laboratory for Laser Energetics, 250 East River Road, Rochester, NY 14623-1299, USA}

\author{D. Haberberger}
\affiliation{University of Rochester Laboratory for Laser Energetics, 250 East River Road, Rochester, NY 14623-1299, USA}

\author{S. T. Ivancic}
\affiliation{University of Rochester Laboratory for Laser Energetics, 250 East River Road, Rochester, NY 14623-1299, USA}

\author{C. A. Walsh}
\affiliation{Lawrence Livermore National Laboratory, Livermore, CA, 94550, USA}

\author{J. R. Davies}
\affiliation{University of Rochester Laboratory for Laser Energetics, 250 East River Road, Rochester, NY 14623-1299, USA}


\date{\today}

\begin{abstract}
Angular filter refractometry is an optical diagnostic that measures absolute contours of line-integrated density gradient by placing a filter with alternating opaque and transparent zones in the focal plane of a probe beam, which produce corresponding alternating light and dark regions in the image plane. Identifying transitions between these regions with specific zones on the angular filter (AF) allows the line-integrated density to be determined, but the sign of the density gradient at each transition is degenerate and must be broken using other information about the object plasma. Additional features from diffraction in the filter plane often complicate data analysis. In this paper, we present an improved AF design that uses a stochastic pixel pattern with a sinusoidal radial profile to minimize unwanted diffraction effects in the image caused by the sharp edges of the filter bands. We also present a technique in which a pair of AFs with different patterns on two branches of the same probe beam can be used to break the density gradient degeneracy. Both techniques are demonstrated using a synthetic diagnostic and data collected on the OMEGA EP laser.
\end{abstract}


\maketitle 

\section{Introduction}

\begin{figure*}
\includegraphics[width=\textwidth]{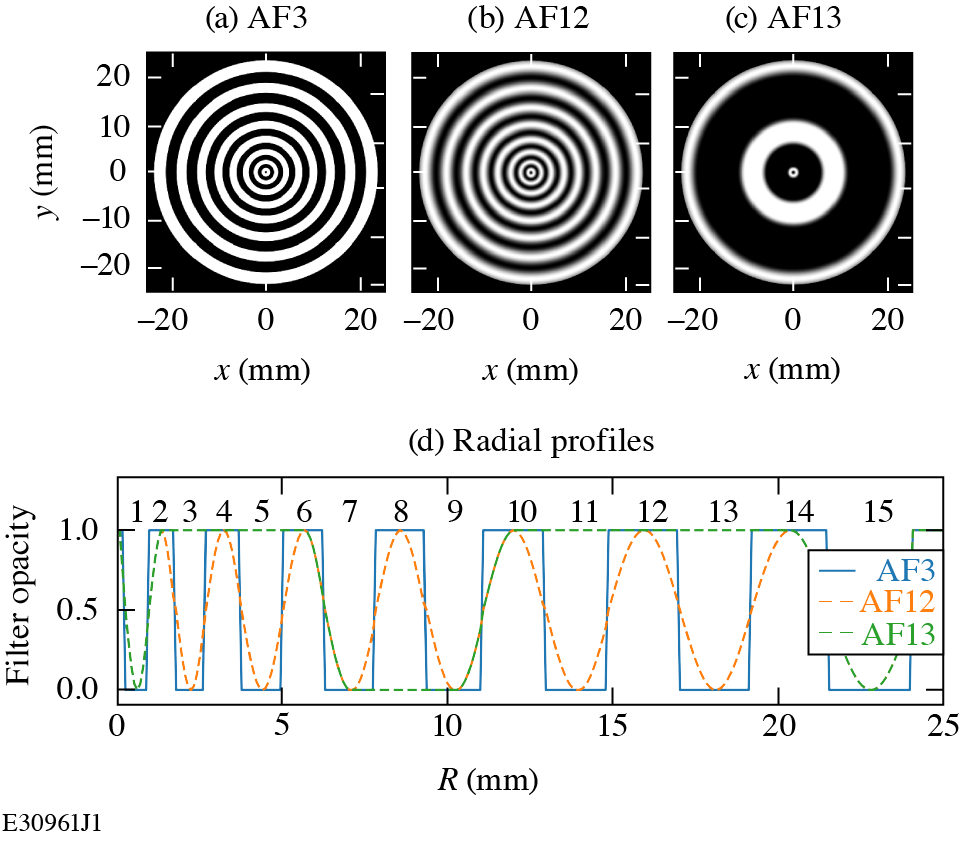}
\caption{Two-dimensional [a--c] and [d] radial 1-D profiles for each of the three AFs discussed. AF3 is an existing OMEGA EP AF, while AF12 and AF13 were fabricated for this experiment. The filter names are assigned chronologically as new filters are built for OMEGA EP. The numbers above the profile in (d) label each filter zone for reference.\label{filter_comparison}}
\end{figure*}

Angular filter refractometry (AFR) is an optical diagnostic that measures absolute contours of line-integrated electron density gradient from which, with some assumptions, the line-integrated electron density profile can be inferred~\cite{Haberberger2014measurements,Kogelschatz1972quantitative}. A probe beam passes through an object plasma, where it is refracted. The beam is collected and collimated by a pair of lenses, then imaged by a third lens onto a detector. An angular filter (AF) comprising a ``bull's eye" pattern of concentric opaque and transparent zones [Figs.~\ref{filter_comparison}(a--c)] is placed in the focal plane of the imaging lens. The center of the filter is opaque, meaning that rays that are not deflected by the object plasma are blocked by the AF. Light that is refracted into the transparent zones on the AF is transmitted to the detector, forming bands of constant refraction angle, or equivalently,  of constant line-integrated density. 

A ray originally parallel to $\hat z$ refracted in the $x$ direction by an electron density in the object plane $n_\text{e}(x_\text{o}, y_\text{o})$ will pass through the filter plane at a radius~\cite{Haberberger2014measurements}
\begin{equation}
r = \frac{1}{2 \xi n_\text{c}} \frac{\partial}{\partial x_\text{o}} \int n_\text{e} dz_\text{o} \text{,}
\end{equation}
%
%
where $n_\text{c}$ is the plasma critical density at the wavelength of the probe beam and $\xi$ is an empirically determined constant determined by the optical geometry~\cite{Haberberger2014measurements}. If the final radial position in the filter plane is known for rays passing through each point in the object plane $r(x_\text{o}, y_\text{o})$, then the line-integrated density corresponding to each point is 
\begin{equation}
\int n_\text{e} dz_\text{o} = 2 \xi n_\text{c} \int r(x_\text{o}, y_\text{o}) dx_\text{o} + C \text{,}
\label{density_formula}
\end{equation}
where integration constant $C$, which includes the initial phase, must be determined by absolutely identifying the filter radius corresponding to one contour (usually with reference to a region of presumed zero refraction).

The 4$\omega$ probe diagnostic on the OMEGA EP laser~\cite{Froula2012optical} consists of a single 10-mJ, 10-ps 4$\omega$ (263-nm) probe laser pulse that is divided into four legs after passing through the object plasma. Up to two of these legs can be configured for angular filter refractometry. This instrument is routinely used to diagnose a wide variety of laser-produced plasmas. Data recorded with the existing AFs, however often include many diffraction artifacts, which are caused by diffraction at the sharp edges of the zones in the filter plane. These artifacts are often difficult to distinguish from the AFR contours of interest (especially in regions with high-density gradients, where the light and dark bands are thin), significantly complicating or preventing analysis.

Additionally, AFR inherently gives degenerate results for nonmonotonic density profiles. If a density profile is presumed to be monotonic (e.g., a single expanding laser-produced plasma), then each transition from a light to a dark band in the image can be unambiguously associated with an increase or decrease in refraction angle. However, if the density profile is nonmonotonic (e.g., colliding plasma flows~\cite{Schaeffer2017high}), then a transition from a light to a dark band could correspond to either an increase or decrease in refraction angle. This degeneracy limits the utility of AFR (without additional assumptions) for many experiments.

In this paper we examine one existing AF design and two improved AF designs that address these issues. The three filters discussed are shown in Fig.~\ref{filter_comparison}.  In Section~\ref{sec:synthetic_diagnostic} we develop a synthetic AFR diagnostic to test and illustrate the behavior of these designs. In Sec.~\ref{sec:data:experiment} we present AFR data from an experiment in which all three AFR filter designs were directly compared. Section~\ref{sec:data:synthetic} applies the synthetic AFR diagnostic to simulations of the experiment and an analytic test profile, and Sec.~\ref{sec:data:reconstruction} demonstrates how the density profile can be reconstructed from lineouts taken from the experimental and synthetic data. Section~\ref{sec:diffraction} describes the use of a stochastic sinusoidal pattern to mitigate diffraction artifacts in the two new AFs, while Sec.~\ref{sec:degeneracy} shows how this pair of new AFs can be used to break degeneracy in the sign of a nonmonotonic density gradient. Our conclusions are summarized in Sec.~\ref{sec:conclusion}.

\section{Creating Synthetic AFR Data\label{sec:synthetic_diagnostic}} 

Synthetic AFR images are generated from 3-D arrays of electron density created either by analytic models or simulations. The electron density is line-integrated along one dimension (here assumed without loss of generality to be along the z-axis). The phase shift of the probe beam through the line-integrated density $\int n_\text{e} dz$ is 
\begin{equation}
\Delta \phi (x_\text{o},y_\text{o}) = -\frac{\omega_\text{p}}{2 c n_\text{c}} \int n_\text{e} dz \text{,}
\end{equation}
where $\omega_\text{p}$ is the angular frequency of the probe beam, $n_\text{c}$ is the critical density at $\omega_\text{p}$, and $(x_\text{o}, y_\text{o})$ are the spatial coordinates in the object plane. Neglecting absorption in the object plasma, the amplitude of the transmitted signal is then 
\begin{equation}
I_0(x_\text{o},y_\text{o}) = e^{i \Delta \phi} \text{.}
\end{equation}
This transmitted signal is then propagated to the focal plane of the imaging lens, also known as the Fourier plane because the intensity distribution in the focal plane is the magnitude of the spatial Fourier transform of $I_0(x_\text{o},y_\text{o})$~\cite{Hecht2016optics}
\begin{equation}
I_f(k_x, k_y) = | F[I_0(x_\text{o},y_\text{o})] |^2 \text{,}
\end{equation}
where $F$ is the Fourier transform and $k_x$,$k_y$ are the spatial frequencies. The spatial coordinates of the Fourier transform in the filter plane corresponding to each spatial frequency are
\begin{align}
x_f  &= k_x \lambda_\text{p} f \text{,}\\
y_f  &= k_y \lambda_\text{p} f \text{,}
\end{align}
where $\lambda_\text{p}$ is the probe beam wavelength and $f$ is the focal length of the imaging lens. Once the intensity in the filter plane $I_f(x_f, y_f)$ is known, the intensity is multiplied by the AF opacity mask $M(x_f, y_f)$ to calculate the signal transmitted through the AF. Applying an inverse Fourier transform then propagates the signal to the image plane.
\begin{equation}
I_I(x_\text{i}, y_\text{i}) = F^{-1} 
\bigg [  
M(x_f, y_f) F[ I_0(x_\text{o}, y_\text{o})  ]
\bigg] \text{.}
\end{equation}
As a final step, two masks are applied to the final image. The first accounts for the fact that rays deflected by more than $\sim7^\circ$ will not pass through the collection lens. The deflection angle for each point in the line-integrated density profile is calculated as
\begin{equation}
\nabla \theta = \frac{1}{2 n_\text{c}} \nabla n_\text{e} \text{d}l
\end{equation}
and regions where $|\nabla \theta|>7^\circ$ are set to zero in the image. The second mask accounts for any regions of density greater than the critical density that would absorb the probe beam. Regions of the original 3-D array of electron density where $n_\text{e} > n_\text{c}$ are identified, and rays that pass through these pixels are set to zero in the synthetic AFR data, representing absorption of the probe beam in the object.

\section{Experimental and Synthetic Data\label{sec:data}}

\subsection{Experimental Measurements\label{sec:data:experiment}}
\begin{figure*}
\includegraphics[width=\textwidth]{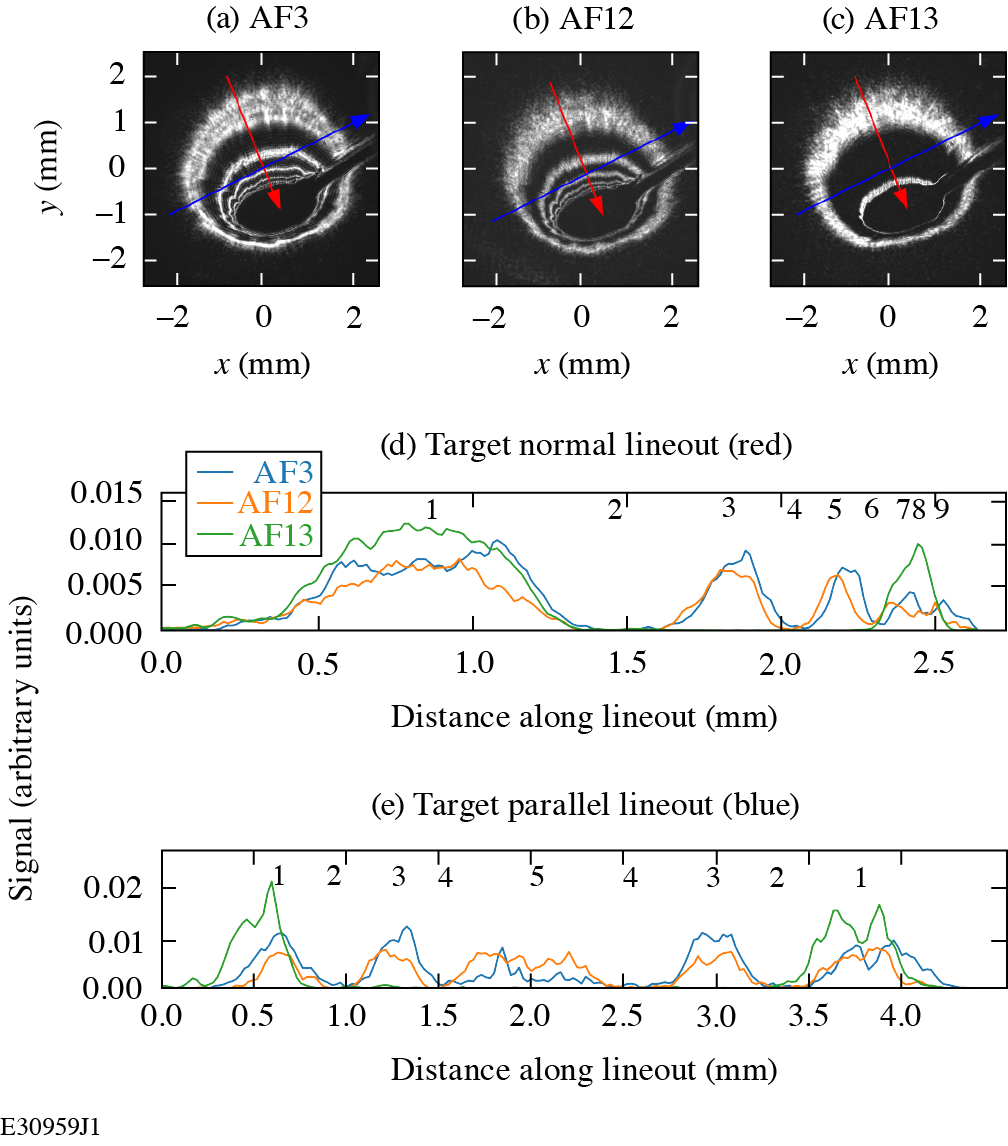}
\caption{(a)-(c) Experimental AFR images with each of the three filters. Images with (a) AF3 and (b) AF12 are taken on the same shot, while (c) the AF13 image is taken on a nominally identical shot. Red and blue arrows indicate the location and direction of lineouts parallel (blue) and normal (red) to the target surface. (d) The perpendicular and (e) parallel lineouts are shown with the inferred filter zones are numbered above each lineout.\label{data_lineouts}}
\end{figure*}

The new AFs described in this paper were tested on an experiment with a simple laser--foil interaction. A 100-$\mu$m-thick 1-mm-outer-diameter silicon disk was positioned approximately edge-on to the 4$\omega$ probe beam on OMEGA EP and illuminated with a 2.2-kJ, 2-ns square pulse UV beam with an $\sim 400$-$\mu$m-diameter spot. AFR images were recorded 2.3~ns after the UV beam on two arms of the 4$\omega$ probe diagnostic with different filters in pairs (AF12+AF13 then AF12+AF3) so that the results could be directly compared. 

Figures~\ref{data_lineouts}(a), \ref{data_lineouts}(b), and \ref{data_lineouts}(c) show data from two shots in this experiment. On the first shot, AF3 and AF12 were fielded simultaneously on two separate legs of the AFR diagnostic, while on a second (nominally identical) shot, AF12 and AF13 were fielded. Images from different legs or shots are aligned using shared features in the image. Lineouts [Figs.~\ref{data_lineouts}(d,e)] are taken by choosing sampling points along the line and interpolating their value from the data.

\subsection{Synthetic Data\label{sec:data:synthetic}}

\begin{figure*}
\includegraphics[width=\textwidth]{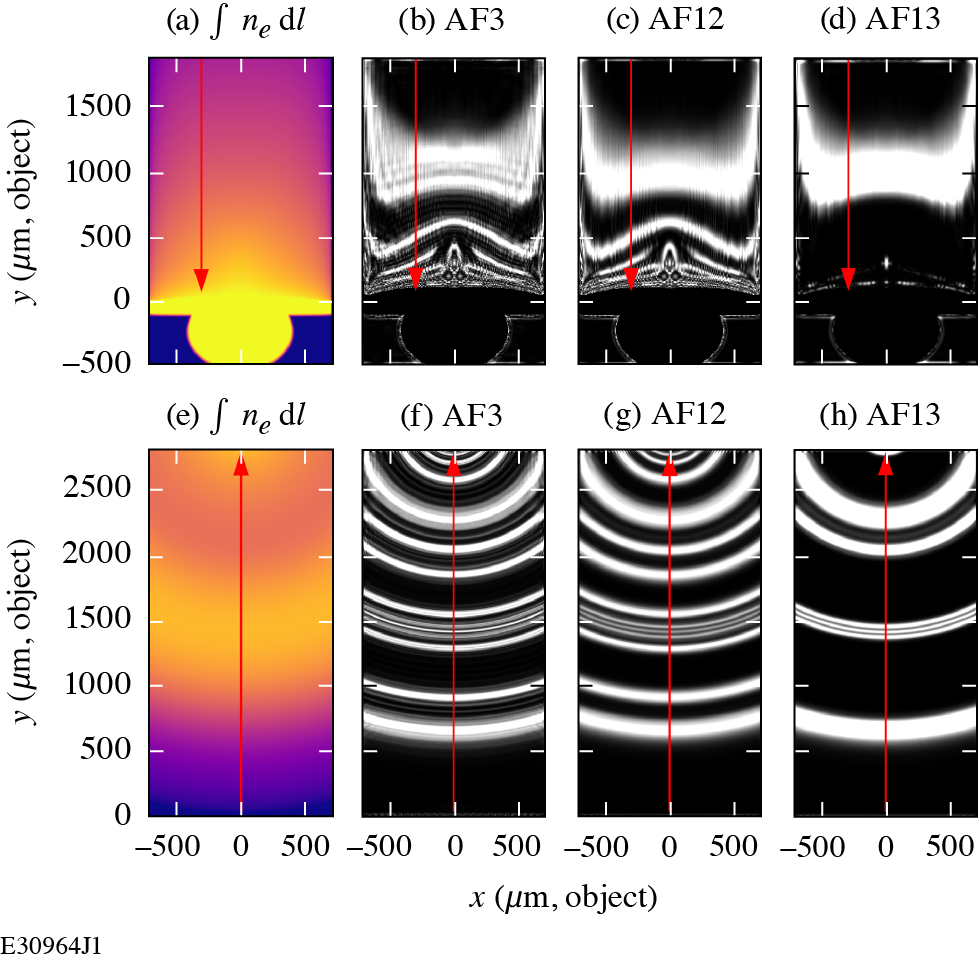}
\caption{ (a) A line-integrated density profile for synthetic dataset 1
generated by a GORGON simulation similar to the experimental setup described in Sec.~\ref{sec:data:experiment}, and [(b)--(d)] corresponding synthetic AFR data using three different AFs. (e) An analytic line-integrated density profile for synthetic dataset 2, and [(f)--(h)] corresponding synthetic AFR data. The red arrows indicate the location and direction of lineouts analyzed in Fig~\ref{lineout_analysis}. Vertical stripes in (b)--(d) are an artifact of limited resolution in the simulation.\label{synthetic_afr}}
\end{figure*}

(a) Synthetic dataset 1 and (e) dataset 2 and corresponding synthetic AFR data [(b)-(d)] and [(f)-(h)].

Figure~\ref{synthetic_afr} shows synthetic AFR images generated from pre-shot simulations (top row, ``synthetic dataset 1") and an analytical model (bottom row, ``synthetic dataset 2"). The pre-shot simulations were performed with the extended-magnetohydrodynamic code GORGON~\cite{Chittenden2004xray,Ciardi2007evolution,Walsh2020extended}, which has previously been applied to similar laser--foil experiments~\cite{Campbell2020magnetic, Campbell2022measuring}. The simulation setup is similar to the experimental setup described previously, with a laser coming from $+\hat y$ incident on a foil at $y=0$. Lineouts taken from the synthetic data are shown in Fig.~\ref{synthetic_afr}. The simulation setup is not perfectly identical to the experiment, resulting in a fractional difference in the simulated density profile and that inferred from the experimental data (Fig.~\ref{lineout_analysis}).  

The second synthetic dataset was created using a nonmonontonic density profile generated by an analytic function of the form
\begin{equation}
n_\text{e} \text{d}l (r) \propto l/r + \frac{3}{4} e^{-\frac{(r-\mu)^2}{\sigma^2}} \text{,}
\end{equation} 
where $r$ is the distance from a central point [0, 3000]~$\mu$m and $l$, $\mu$, and $\sigma$ are constants. The density profile is constant in $z$ over a depth of 1~mm. This produces a nonmonotonic radial density distribution shown in Fig.~\ref{lineout_analysis}(b).

\subsection{Reconstruction Procedure\label{sec:data:reconstruction}}

\begin{figure}
\includegraphics[width=0.5\textwidth]{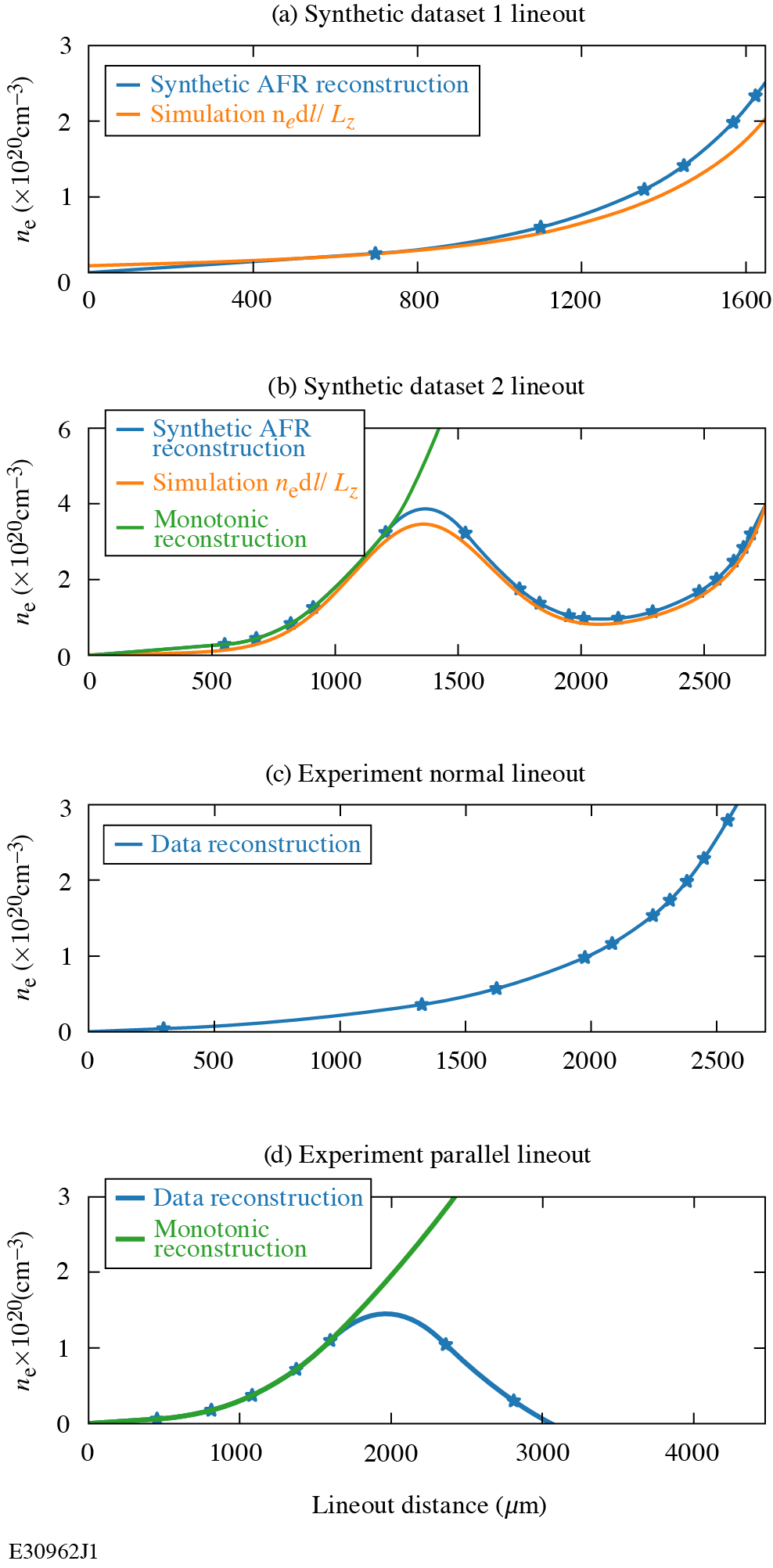}
\caption{Density profiles reconstruction from lineouts taken from AF12 images generated from (a),(b) the  synthetic datasets and from (c),(d) the experimental AF12 data. In (a) and (b), the reconstructed density (blue) is compared to the actual line-integrated density (orange). In (b) and (d), the correct analysis using the pair of AF12/AF13 (blue) is compared to the monotonic profile that would be inferred using only AF12 (green). Markers on the blue line in each plot show the location of each filter zone transition used in the reconstruction.  \label{lineout_analysis}}
\end{figure}

Density profiles (Fig.~\ref{lineout_analysis}) are reconstructed from each of the experimental and synthetic lineouts using the following procedure. First the filter zone corresponding to one band in the image is unambiguously identified. For both the synthetic and experimental data, the lineouts start at very low density (zone 0). The edges between the dark and light bands of the lineout are then marked, and the refraction angle corresponding to each edge is determined using the geometry constant $\xi$. The resulting profile is then interpolated and numerically integrated to recover the line-integrated density profile using Eq.~\ref{density_formula}. The analysis of the synthetic datasets reproduces the ground truth density profile [Fig.~\ref{lineout_analysis}(a)-(b)]. Unambiguous analysis of the nonmonotonic density profile in synthetic dataset 2 [Fig.~\ref{lineout_analysis}(b)] and the target-parallel lineout [Fig.~\ref{lineout_analysis}(d)] requires the use of a pair of AFR filters, as described in Sec.~\ref{sec:degeneracy}.

\section{Mitigating Diffraction Artifacts\label{sec:diffraction}}

\begin{figure}
\includegraphics[width=0.5\textwidth]{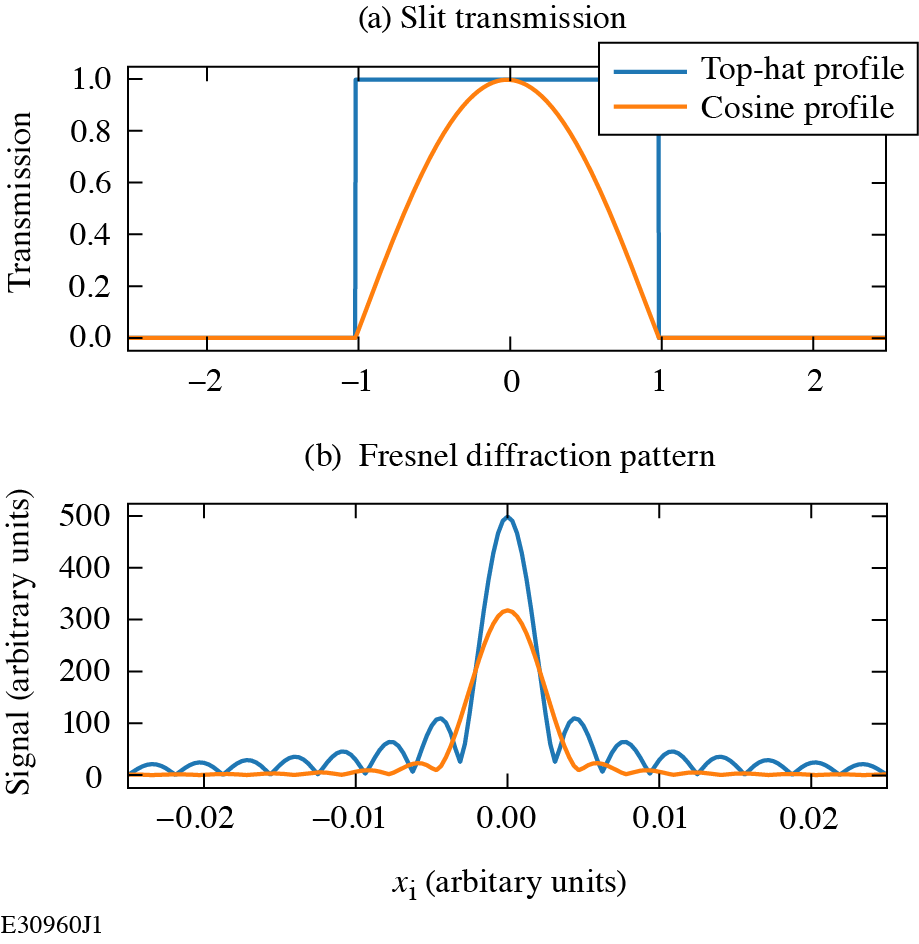}
\caption{Fresnel diffraction through a slit with a cosine transmission profile produces significantly reduced higher-order peaks than diffraction through a slit with a top-hat profile.\label{diffraction_example}}
\end{figure}

\begin{figure}
\includegraphics[width=0.5\textwidth]{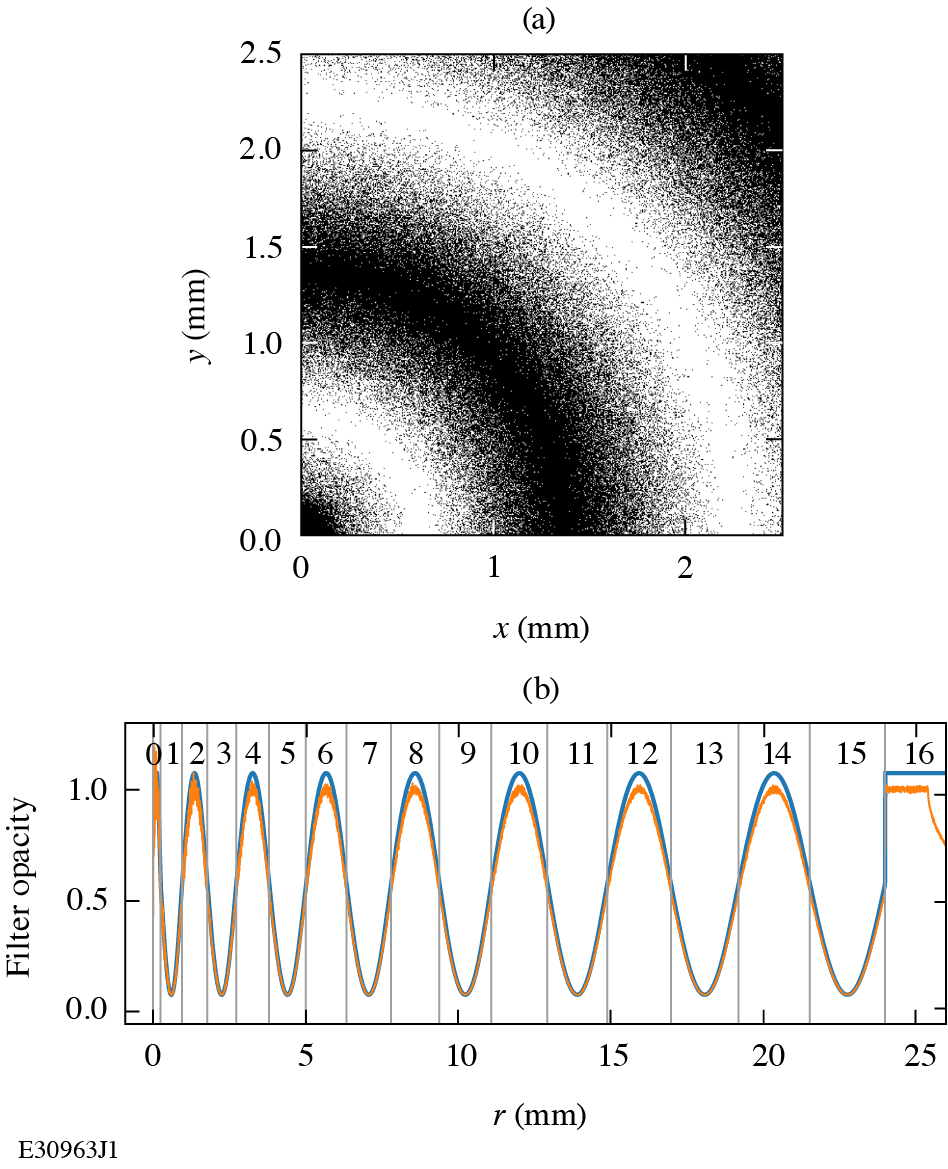}
\caption{(a) A closeup of the pixel values for AF12 showing the stochastic distribution of binary transmission values. (b) The nominal radial transmission profile of AF12 (blue) compared with a radial histogram of actual stochastic pixel values (orange) shows that the sochastic process produces the desired transmission profile when averaged around the filter. The histogram is multiplied by $1/r$ for comparison to the nominal radial profile.\label{stochastic_filter}}
\end{figure}

Previous AFs fielded on OMEGA EP employed a top-hat transmission profile [e.g., AF3, Fig.~\ref{filter_comparison}(a)], which produces many higher-order diffraction peaks. Depending on their prominence and width relative to other features, these peaks can easily be mistaken for additional contours and counted during analysis, leading to unphysical reconstructed density profiles. These higher-order diffraction peaks can be significantly suppressed by using an AF with a sinusoidal radial transmission profile (Fig.~\ref{diffraction_example}) such as the ones shown in Fig.~\ref{filter_comparison}(d). The effect of this sinusoidal profile is visible in the synthetic images as smoother zone structure with the sinusoidal filter AF12 [Fig.~\ref{synthetic_afr}(c)] compared to the top-hat filter AF3 [Fig.~\ref{synthetic_afr}(b)].

The AFs used on OMEGA EP are produced by depositing small regions of metal on a glass substrate. This method allows the transmission value of each pixel to be individual specified. It means, however, that AF pixel values are inherently binary, making a truly continuous transmission profile unachievable with this technique. 
 
For AF12 and AF13, a cosine transmission profile is instead approximated by randomly drawing binary pixel values from a sinusoidal radial probability distribution [Fig.~\ref{stochastic_filter}(a)]. Each pixel is $5$ $\mu$m $\times$ $5$ $\mu$m, and each filter is defined by specifying an array of $10160$ $\times$ $10160$ pixels. The radial distance of each pixel's center from the center of the array is calculated, then used to interpolate the nominal filter transmission at that point from the curves shown in Fig.~\ref{filter_comparison}(d). These nominal values are then treated as a probability distribution from which the actual binary pixel values are drawn. Since the filter comprises $\sim100$ million pixels, this operation is computed in parallel. When averaged over the polar angle of the filter, this produces the desired sinusoidal transmission profile as shown in Fig.~\ref{stochastic_filter}(b). Due to the number of pixels in the stochastic filter array, a continuous (nonbinary) radial profile is used for the synthetic data, rather than the actual filter pixel values. 

Experimental data confirm that the stochastic sinusoidal filter, AF12, produces fewer diffraction artifacts than the otherwise identical top-hat filter AF3. For example the extra bands visible in Fig.~\ref{data_lineouts}(a) are diffraction artifacts that have been mitigated by AF12 in Fig.~\ref{data_lineouts}(b). Some small higher-order peaks are visible in the AF3 lineout in Fig.~\ref{data_lineouts}(d) that are smoothed out in the orange AF12 lineout. 

\section{Nondegenerate Density-Gradient Measurements and Complementary Filters\label{sec:degeneracy}} 

The procedure for reconstructing a density profile from an AFR image described in Sec.~\ref{sec:data:reconstruction} relies on an implicit assumption that the density profile is monotonic, such that each successive dark or light band in the image always represents the next-higher band on the filter, rather than the next-lower band. Nonmonotonic density profiles, however are common, e.g., colliding plasma flows~\cite{Schaeffer2017high} or density cavities. Synthetic dataset 2 illustrates such a nonmonotonic density profile, and the green curve in Fig.~\ref{lineout_analysis}(b) shows how an unphysical density profile will be recovered if the profile is assumed to be monotonic. A second trivial example is given by the experimental lineouts parallel to the target surface in Fig.~\ref{data_lineouts} which cross the symmetry axis of the expanding laser-produced plasma and therefore have a nonmonotonic density profile. 

This degeneracy can be broken by using a pair of AFs with complementary filter patterns (AF12 and AF13). Consider an example in which a bright band has been unambiguously identified as corresponding to zone 9 on AF12 [see Fig.~\ref{filter_comparison}d]. A subsequent dark band could correspond to either an increasing refraction (zone 10) or a decreasing refraction (zone 8). However, comparison with an image from AF13 breaks this degeneracy, since only zone 10 is also opaque on AF13. This method was applied to the experimental data to identify the bands in Fig.~\ref{data_lineouts}(c), and to subsequently recover the correct density profile in Figs.~\ref{lineout_analysis}(b) and \ref{lineout_analysis}(d). 

In principle, the density profile can remain ambiguous with two complementary filters if the density jumps suddenly by more than one filter zone. This is possible at sharp discontinuities such as shocks. In this case, additional simultaneous filters would be required to resolve the degeneracy. The complementary filter presented here (AF13) is also not designed to detect reversals in the density gradient corresponding to deflections of less than two zones. For example, a pattern of deflections from zones $3 \rightarrow 4 \rightarrow 3$ would be indistinguishable from the pattern $3 \rightarrow 4 \rightarrow 5$. This ambiguity could be resolved on future complementary filters, but at the cost of potentially introducing ambiguity at large changes in density. Fielding multiple complementary filters (three or more AFs total) could simultaneously resolve both ambiguities.  

We have also considered the possibility of several other complementary sets of AFs. If a given experimental setup does not have a region of known zero density (or near-zero density gradient), an AF with only a single transparent band could be used to provide an unambiguous reference. Alternatively, multiple AFs with progressively larger band spacing could be used to achieve higher-dynamic-range measurements. We leave implementation and testing of these designs to future work. 

\section{Conclusion\label{sec:conclusion}}

A set of two new AF designs was tested using a synthetic AFR diagnostic and an experiment on the OMEGA EP laser. The new designs utilize a stochastic pattern of binary value pixels that, when averaged in the polar angle of the filter, creates a sinusoidal opacity profile that minimizes diffraction artifacts in the images. This new filter design is successful in minimizing diffraction artifacts in both synthetic and experimental measurements when compared to an existing AF design with a top-hat radial opacity profile. A second filter, identical to the first except with an alternating pattern of opaque bands, is used to unambiguously reconstruct nonmonotonic density gradients. Several additional possible designs for complementary sets of AFs that could provide absolute reference bands or achieve higher dynamic gain are proposed but left for future work.

\begin{acknowledgments}
This material is based upon work supported by the Department of Energy National Nuclear Security Administration under Award Numbers DE-NA0003856 and DE-SC0020431, the University of Rochester, and the New York State Energy Research and Development Authority. 

This report was prepared as an account of work sponsored by an agency of the U.S. Government. Neither the U.S. Government nor any agency thereof, nor any of their employees, makes any warranty, express or implied, or assumes any legal liability or responsibility for the accuracy, completeness, or usefulness of any information, apparatus, product, or process disclosed, or represents that its use would not infringe privately owned rights. Reference herein to any specific commercial product, process, or service by trade name, trademark, manufacturer, or otherwise does not necessarily constitute or imply its endorsement, recommendation, or favoring by the U.S. Government or any agency thereof. The views and opinions of authors expressed herein do not necessarily state or reflect those of the U.S. Government or any agency thereof.
\end{acknowledgments}


%

\end{document}